\documentclass[showpacs,showkeys,twocolumn,pra]{revtex4}%
\usepackage{graphics,amsmath,amsfonts,amscd,revsymb,latexsym,enumerate,multirow,epsfig}
\usepackage{amsmath}
\usepackage{amsfonts}
\usepackage{amssymb}
\usepackage{graphicx}%
\providecommand{\U}[1]{\protect\rule{.1in}{.1in}}
\providecommand{\U}[1]{\protect\rule{.1in}{.1in}}

\newcommand{\qed}{{\hfill$\Box$}}

\def\bi{\begin{itemize}}
\def\ei{\end{itemize}}
\def\be{\begin{equation}}
\def\ee{\end{equation}}
\def\bea{\begin{eqnarray}}
\def\eea{\end{eqnarray}}
\def\ben{\begin{eqnarray*}}
\def\een{\end{eqnarray*}}

\def\>{\rangle}
\def\<{\langle}

\newcommand{\1} I

\def\*{\star}

\def\0{{\mathbf{0}}}
\def\1{{\mathbf{1}}}
\def\2{{\mathbf{2}}}
\def\3{{\mathbf{3}}}
\def\4{{\mathbf{4}}}
\def\5{{\mathbf{5}}}
\def\6{{\mathbf{6}}}
\def\7{{\mathbf{7}}}
\def\8{{\mathbf{8}}}
\def\9{{\mathbf{9}}}

\begin{document}
\title{Could light harvesting complexes exhibit non-classical effects at room temperature?}
\author{Mark M. Wilde}
\author{James M. McCracken}
\affiliation{Electronic Systems Division, Science Applications International Corporation,
4001 North Fairfax Drive, Arlington, Virginia, USA\ 22203}
\author{Ari Mizel}
\affiliation{Laboratory for Physical Sciences, 8050 Greenmead Drive, College Park,
Maryland, USA\ 20740}
\keywords{Leggett-Garg inequality, macrorealism, Fenna-Matthews-Olson protein complex,
Prosthecochloris aestuarii, quantum effects in biological systems}\date{\today }

\pacs{03.65.Ta, 03.65.Yz, 82.39.Jn}

\begin{abstract}
Mounting experimental and theoretical evidence suggests that coherent quantum
effects play a role in the efficient transfer of an excitation from a
chlorosome antenna to a reaction center in the Fenna-Matthews-Olson protein
complex. However, it is conceivable that a satisfying alternate interpretation
of the results is possible in terms of a classical theory. To address this
possibility, we consider a class of classical theories satisfying the minimal
postulates of macrorealism and frame Leggett-Garg-type tests that could rule
them out. Our numerical simulations indicate that even in the presence of
decoherence, several tests could exhibit the required violations of the
Leggett-Garg inequality. Remarkably, some violations persist even at room
temperature for our decoherence model.

\end{abstract}
\maketitle

\section{Introduction}

Does quantum mechanics play a nontrivial role in evolutionary or molecular
processes in biological systems? How could this result be true, when
biological systems interact in environments that are \textquotedblleft
hot\textquotedblright\ and \textquotedblleft wet?\textquotedblright
Furthermore, how can we frame the first question \textit{rigorously}?\ These
questions have been the subject of intense debate \cite{AGDHZEWB08,ADP08book}
and a surge of recent interest\ from both the quantum information community
\cite{BP08,CGB09,CPB08,CCDHP09,K08a,K08b,K08c,Lloyd:2009:164,mohseni:174106,1367-2630-10-11-113019,1367-2630-11-3-033003,RGMBV09,Gilmore:2008:2162,olaya-castro:085115}%
\ and the experimental quantum chemistry community
\cite{fleming2007nature,fleming2007science,RTPWW04,CF09}. This active area is
sometimes described as \textquotedblleft quantum biology\textquotedblright%
\cite{ADP08book,Lloyd:2009:164}. Some have suggested that, in addition to the
possibility of remarkable new insights into biology, there is the potential to
harness naturally occurring coherent quantum effects in biology to engineer
new devices.

Two biological processes are playing a particularly important role in fueling
interest in this subject:\ energetic excitation transfer in the
Fenna-Matthews-Olson (FMO) light-harvesting protein complex of green bacteria
\cite{Adolphs20062778,AVG00,CCDHP09,CKPS97,CF09,fleming2007nature,FNO09,Fenna:1975:573,Gilmore:2008:2162,fleming2007science,L96,mohseni:174106,olaya-castro:085115,1367-2630-10-11-113019,RCA09,1367-2630-11-3-033003,Renger2001137}
and the radical ion-pair mechanism in the so-called avian compass
\cite{CGB09,RGMBV09,K08a,K08b,K08c}. For each system, there are claims of
quantum mechanical behavior in the scientific literature. On the theory side,
several studies in the literature have proposed an open quantum systems
approach to model the dynamics of these biological processes
\cite{1367-2630-11-3-033003,mohseni:174106,1367-2630-10-11-113019,CCDHP09} and
have lent credence to the proposition that coherent quantum effects play a
role in their functionality.

Agreement between quantum theoretical models and experiment, however, does not
\textit{irrevocably} demonstrate the presence of quantum effects. This point
is subtle, but it is a logical fallacy, called the \textquotedblleft
affirmation of the consequent,\textquotedblright\ to conclude once and for all
that the quantum biological models are the correct models simply because they
coincide with the observations of some experiments (Leggett stresses this
point in several of his papers \cite{L02,L08}). It would be useful to
\textit{irrevocably} exclude certain classes of classical models that might
apply to these biological systems by considering fundamental tests of non-classicality.

One might wonder if it is well-motivated to conduct a test for
non-classicality on a system such as the FMO\ complex that is clearly
microscopic. In fact, is it not obvious that the FMO\ complex should behave
according to quantum-mechanical laws, given that the length scales and time
scales are those to which quantum theory applies?\ We argue that applying a
test for non-classicality is valuable here because a significant amount of
environmental noise acts on the chromophores of the FMO\ complex. These
decohering effects may \textquotedblleft wash out\textquotedblright%
\ quantum-mechanical behavior and make it appear as if the FMO\ complex
behaves according to a class of macrorealistic models (after all, some argue
that classical behavior in general arises due to such decoherence \cite{Z91}).
In some cases (in a high temperature limit), researchers actually have modeled
the dynamics of the FMO\ complex with a classical incoherent hopping
(F\"{o}rster)\ model \cite{F46,KR82,L96,CKPS97}. Recent results
\cite{mohseni:174106,1367-2630-11-3-033003,1367-2630-10-11-113019,CCDHP09}%
\ suggest that a classical incoherent hopping model is insufficient to explain
the ultra-efficient transfer of energy in photosynthesis, but a test for
non-classicality could irrevocably exclude the whole class of models to which
the classical incoherent hopping model belongs.

In this paper, we frame a test for non-classicality in the FMO\ protein
complex, the molecular complex responsible for the transfer of energetic
excitations in a photosynthetic reaction. Quantum chemists have determined a
tight-binding Hamiltonian for this simple system \cite{Adolphs20062778}.
Phenomenological modifications to the standard Schr\"{o}dinger equation have
allowed an open quantum systems model of its dynamics
\cite{1367-2630-11-3-033003,mohseni:174106,1367-2630-10-11-113019,CCDHP09}.
Motivated by experimental results, several theoretical studies have computed
the efficiency of energy transfer from the chlorosome antenna to the reaction
center in green bacteria and have asserted that coherent quantum effects play
a role. To assess this claim, we frame a test for violation of
\textit{macrorealism}, as quantified by the Leggett-Garg inequality
\cite{LG85}. In this sense, we are following the program of Leggett outlined
in Ref.~\cite{L02} and the suggestion of Zeilinger in Ref.~\cite{AGDHZEWB08}%
\ that it would be useful to subject biomolecules to tests of
non-classicality. The Bell \cite{B64,B87}\ and Leggett-Garg \footnote{Leggett
points out \cite{L02} that single-system Bell-type inequalities \cite{LG85}
appeared earlier in Ref. \cite{Home1984159}.}\ tests of non-classicality are
benchmarks that determine whether a given dynamical system has
stronger-than-classical spatial or temporal correlations, respectively. Each
of these tests provides an inequality that bounds the spatial or temporal
correlations of a given system---a violation of the inequality implies that
the system in question does not behave in a classical manner. The advantage of
the Leggett-Garg test over a Bell test is that a Leggett-Garg test applies to
a \textit{single} quantum system, easing experimental difficulty. The
Leggett-Garg inequality was originally applied to superconducting quantum
systems. It was later considered for other systems such as quantum dots
\cite{ruskov:06,jordan:026805,williams:026804} and photons \cite{GABLOWP09}
but, to our knowledge, this is the first application of the Leggett-Garg
theory to a biological system.

We organize the rest of this paper as follows. We introduce the Leggett-Garg
inequality in the next section and review a simple example of a two-level
quantum system that violates it. In Section~\ref{sec:FMO}, we review the
quantum dynamical model of the FMO\ complex (we specifically employ the
nine-level model of Refs.~\cite{CCDHP09,1367-2630-10-11-113019}, which is a
modification of the seven-level model of Rebentrost \textit{et al}. in
Refs.~\cite{olaya-castro:085115,mohseni:174106,1367-2630-11-3-033003}). We
then discuss several examples of dichotomic observables for the FMO\ complex
to which we can apply the Leggett-Garg inequality. Section~\ref{sec:LG-FMO}
begins our study of the Leggett-Garg inequality and the FMO\ complex. We first
study the Leggett-Garg inequality with purely coherent dynamics and are able
to derive analytical results. These analytical formulas allow us to determine
exactly when coherent dynamics give a violation of the inequality. We then
exploit these analytical results in our numerical simulations of the
Leggett-Garg inequality and the FMO\ complex, where we show that it is still
possible to violate the inequality even in the presence of noise and
potentially even at room temperature. We conclude with some open questions for
further study.

\section{The Leggett-Garg Inequality}

\label{sec:LG}The Leggett-Garg inequality applies to any system that obeys the
postulates of macrorealism. The postulates of macrorealism for a two-level
system are as follows \cite{LG85,L02}:

\begin{enumerate}
\item \textit{Macrorealism per se}: A macroscopic object is in one of two
definite states at any given time.

\item \textit{Noninvasive measurement}:\ It is possible in principle to
determine the state of the system without affecting it or any subsequent dynamics.

\item \textit{Induction}: The properties of ensembles are determined
exclusively by initial conditions (and, in particular, not by final conditions).
\end{enumerate}

It is reasonable to assume that a classical system, in principle, should obey
the postulates of a macrorealistic theory.

The Leggett-Garg inequality bounds the two-time correlation functions of three
dichotomic observables $Q\left(  t_{1}\right)  $, $Q\left(  t_{2}\right)  $,
and $Q\left(  t_{3}\right)  $ measured at respective times $t_{1}$, $t_{2}$,
and $t_{3}$. The observables $Q\left(  t_{1}\right)  $, $Q\left(
t_{2}\right)  $, and $Q\left(  t_{3}\right)  $ could be the spin of a particle
or the location of the trapped magnetic flux in a double-well potential as in
the original application of Leggett and Garg \cite{LG85}. Let $C_{i,j}$ denote
the following two-time correlation function:%
\[
C_{i,j}\equiv\left\langle Q\left(  t_{i}\right)  Q\left(  t_{j}\right)
\right\rangle .
\]
The Leggett-Garg quantity $K$ is the following combination of three two-time
correlations and a constant:%
\begin{equation}
K\equiv C_{1,2}+C_{2,3}+C_{1,3}+1. \label{eq:LG-quantity}%
\end{equation}
Note that we can obtain the alternate Leggett-Garg quantities
\begin{align}
&  -C_{1,2}+C_{2,3}-C_{1,3}+1,\label{eq:LG-flip-1}\\
&  -C_{1,2}-C_{2,3}+C_{1,3}+1,\label{eq:LG-flip-2}\\
&  C_{1,2}-C_{2,3}-C_{1,3}+1, \label{eq:LG-flip-3}%
\end{align}
merely\ by flipping the sign of the respective observables $Q\left(
t_{1}\right)  $, $Q\left(  t_{2}\right)  $, and $Q\left(  t_{3}\right)  $. The
following Leggett-Garg inequality bounds the Leggett-Garg quantity $K$ when
the system in question is a macrorealistic system being measured
noninvasively:%
\begin{equation}
K\geq0. \label{eq:LG-inequality}%
\end{equation}
The last correlation function $C_{1,3}$ is to be obtained experimentally by
measuring at times $t_{1}$ and $t_{3}$ but refraining from measuring at time
$t_{2}$. By comparing $C_{1,3}$ to correlation functions $C_{1,2}$ and
$C_{2,3}$ obtained in the presence of a measurement at $t_{2}$, the
Leggett-Garg inequality is sensitive to invasiveness in the $t_{2}$ measurement.

An example of a system that violates the Leggett-Garg inequality is a spin-1/2
particle with system Hamiltonian $H=\omega\sigma_{X}/2$ and with the
observables $Q\left(  t_{1}\right)  =Q\left(  t_{2}\right)  =Q\left(
t_{3}\right)  =\sigma_{Z}$. This choice leads to the following value of the
Leggett-Garg quantity $K$ in (\ref{eq:LG-quantity}):%
\[
K=\cos\left(  2\omega\Delta t\right)  +2\cos\left(  \omega\Delta t\right)
+1,
\]
where the parameter $\Delta t$ is the uniform time interval between the
successive measurements of the observable $Q$. Observe that choosing the
interval $\Delta t=3\pi/4\omega$ sets $K=-\sqrt{2}+1$ and leads to a violation
of the Leggett-Garg inequality in (\ref{eq:LG-inequality}). Thus, this quantum
system does not obey the postulates of a macrorealistic theory when we choose
the measurement time intervals as given above. This violation is perhaps not
surprising because a spin-$1/2$ system is a genuine quantum system and
\textquotedblleft cannot have the objective properties tentatively attributed
to macroscopic objects prior to and independent of
measurements\textquotedblright\ \cite{P95}.

\section{Model for the FMO Complex}

\label{sec:FMO}Much of the \textquotedblleft quantum
biological\textquotedblright\ interest has focused on energy transport in the
Fenna-Matthews-Olson (FMO)\ protein complex \cite{Fenna:1975:573}, which is
believed to be the main contributor to ultra-efficient energy transfer in
photosynthesis. The FMO protein complex is a trimer in the bacterial species
\textit{prosthecochloris aestuarii}. The theoretical models in the literature
\cite{CCDHP09,1367-2630-10-11-113019,olaya-castro:085115,mohseni:174106,1367-2630-11-3-033003}%
\ apply to the dynamics of one unit of the trimer. The model assumes that a
photon impinges on the peripheral antenna of the light harvesting complex.
Absorption of the photon produces an electronic excitation, an exciton, that
then traverses a network of seven chromophores or \textit{sites} in one unit
of the trimer. The exciton can either recombine, representing a loss of the
excitation, or it can transfer to a reaction center, where a light-to-charge
conversion occurs before energy storage. Theoretical models
\cite{mohseni:174106,1367-2630-11-3-033003} indicate that coherent quantum
effects combined with decoherence may lead to a quantum stochastic walk
\cite{RWA09} that transports energy efficiently. Rebentrost \textit{et al}.
provide evidence that coherent quantum effects are responsible for the
ultra-high efficiency of photosynthesis \cite{1367-2630-11-3-033003} by
demonstrating that the transport efficiency is much higher with coherent
quantum effects than it is without.

Our physical model for excitation transfer is the nine-level model in
Refs.~\cite{CCDHP09,1367-2630-10-11-113019}, a modification of the seven-level
model in Refs.~\cite{olaya-castro:085115,mohseni:174106,1367-2630-11-3-033003}%
. We can restrict dynamics to the single-excitation manifold because the
excitation number is a conserved quantity in the absence of light-matter
interaction events (within the exciton recombination time scale of 1 ns
\cite{TGOwens031987}).
The possible states for the exciton can be expressed in the \textit{site
basis} $\left\{  \left\vert m\right\rangle \right\}  _{m=1}^{7}$, where the
state $\left\vert m\right\rangle $ indicates that the excitation is present at
site $m$. The incoherent dynamics include a \textquotedblleft
ground\textquotedblright\ state $\left\vert G\right\rangle $ corresponding to
the loss or recombination of the excitation and a sink state $\left\vert
S\right\rangle $ corresponding to the trapping of the exciton at the reaction
center. The excitation evolves into one of the two states $\left\vert
G\right\rangle $ or $\left\vert S\right\rangle $\ in the limit of infinite
time. The density operator $\rho$\ for this open quantum system admits the
following representation in the site\ basis:%
\[
\rho=\sum_{m,n\in\left\{  G,1,\ldots,7,S\right\}  }\rho_{m,n}\left\vert
m\right\rangle \left\langle n\right\vert .
\]

We simplify our analysis by assuming that the dynamics of the density operator
are Markovian. Thus, a Lindblad master equation, with coherent and incoherent
components, models the dynamics \cite{L76,BP07}. Coherent evolution occurs
according to the following Hamiltonian $H$ \cite{Adolphs20062778}:%
\begin{multline}
H\equiv\label{eq:site-hamil}\\%
\begin{bmatrix}
215 & -104.1 & 5.1 & -4.3 & 4.7 & -15.1 & -7.8\\
-104.1 & 220 & 32.6 & 7.1 & 5.4 & 8.3 & 0.8\\
5.1 & 32.6 & 0 & -46.8 & 1.0 & -8.1 & 5.1\\
-4.3 & 7.1 & -46.8 & 125 & -70.7 & -14.7 & -61.5\\
4.7 & 5.4 & 1.0 & -70.7 & 450 & 89.7 & -2.5\\
-15.1 & 8.3 & -8.1 & -14.7 & 89.7 & 330 & 32.7\\
-7.8 & 0.8 & 5.1 & -61.5 & -2.5 & 32.7 & 280
\end{bmatrix}
,
\end{multline}
where the above matrix representation of $H$ is with respect to the site basis
$\left\{  \left\vert m\right\rangle \right\}  _{m=1}^{7}$, and the units of
energy are cm$^{-1}$ (the typical units of choice in spectroscopy
experiments). The diagonal terms in $H$\ correspond to the site energies, and
the off-diagonal terms correspond to intersite couplings. The order of
magnitude of the energies in the above Hamiltonian is 100 cm$^{-1}$, implying
that we should observe dynamics on the order of 300 femtoseconds (fs) (Fleming et al.
experimentally observed behavior on this order in Ref.~\cite{fleming2007nature}).

Three Lindblad superoperators \cite{L76,BP07}\ also contribute to the dynamics
of the density operator in the nine-level model in
Refs.~\cite{CCDHP09,1367-2630-10-11-113019}. The general form of a Lindblad
superoperator $\mathcal{L}\left(  \rho\right)  $ is
\[
\mathcal{L}\left(  \rho\right)  \equiv\sum_{m}\zeta_{m}\left(  2A_{m}\rho
A_{m}^{\dag}-\left\{  A_{m}^{\dag}A_{m},\rho\right\}  \right)  ,
\]
where $\zeta_{m}$ is a rate and $A_{m}$ is a Lindblad operator \cite{L76,BP07}.

The first Lindblad superoperator $\mathcal{L}_{\text{diss}}$ in our model
corresponds to the dissipative recombination of the exciton (loss of energy in
the system):%
\begin{equation}
\mathcal{L}_{\text{diss}}\left(  \rho\right)  \equiv\sum_{m=1}^{7}\Gamma
_{m}\left(  2\left\vert G\right\rangle \left\langle m\right\vert
\rho\left\vert m\right\rangle \left\langle G\right\vert -\left\{  \left\vert
m\right\rangle \left\langle m\right\vert ,\rho\right\}  \right)  .
\label{eq:lindblad-diss}%
\end{equation}
An excitation at site $\left\vert m\right\rangle $ recombines with rate
$\Gamma_{m}$, and $\left\vert G\right\rangle \left\langle m\right\vert $ is
the Lindblad operator that effects this dissipation.

The next Lindblad superoperator $\mathcal{L}_{\text{sink}}\left(  \rho\right)
$ accounts for the trapping of the exciton in the reaction center:%
\begin{equation}
\mathcal{L}_{\text{sink}}\left(  \rho\right)  \equiv\Gamma_{\text{sink}%
}\left(  2\left\vert S\right\rangle \left\langle 3\right\vert \rho\left\vert
3\right\rangle \left\langle S\right\vert -\left\{  \left\vert 3\right\rangle
\left\langle 3\right\vert ,\rho\right\}  \right)  . \label{eq:lindblad-sink}%
\end{equation}
The Lindblad superoperator $\mathcal{L}_{\text{sink}}$\ includes the operator
$\left\vert S\right\rangle \left\langle 3\right\vert $ because evidence
suggests that site~3 in the FMO\ complex plays a crucial role in transferring
the exciton to the reaction center \cite{Adolphs20062778}, where it is later
exploited for energy storage.

The final Lindblad superoperator $\mathcal{L}_{\text{deph}}$ accounts for the
unavoidable dephasing interaction with the environment:%
\begin{equation}
\mathcal{L}_{\text{deph}}\left(  \rho\right)  \equiv\sum_{m=1}^{7}\gamma
_{m}\left(  2\left\vert m\right\rangle \left\langle m\right\vert
\rho\left\vert m\right\rangle \left\langle m\right\vert -\left\{  \left\vert
m\right\rangle \left\langle m\right\vert ,\rho\right\}  \right)  ,
\label{eq:lindblad-deph}%
\end{equation}
where $\gamma_{m}$ is the rate of dephasing at site $m$. Discussion of this
rough treatment of the decoherence appears in Refs.
\cite{1367-2630-11-3-033003,Haken:1973:135,L96}.

The following Lindblad quantum master equation governs the evolution of the
density operator $\rho$:%
\begin{equation}
\dot{\rho}=-i\left[  H,\rho\right]  +\mathcal{L}_{\text{diss}}\left(
\rho\right)  +\mathcal{L}_{\text{sink}}\left(  \rho\right)  +\mathcal{L}%
_{\text{deph}}\left(  \rho\right)  , \label{eq:FMO-dynamics}%
\end{equation}
where we explicitly see the contribution of the Hamiltonian
(\ref{eq:site-hamil}) and the noise superoperators (\ref{eq:lindblad-diss}%
-\ref{eq:lindblad-deph}) to the dynamics, and we implicitly set $\hbar=1$.
Evolution according to the above Lindblad evolution equation is
completely-positive and trace-preserving (CPTP) for any time \cite{BP07}, and
we let $\mathcal{N}_{t,t_{0}}\left(  \rho\right)  $\ denote the induced CPTP
map corresponding to the evolution of the density operator $\rho$ from an
initial time $t_{0}$ to some later time $t$.

\section{The Leggett-Garg Inequality and the FMO\ Complex}

\label{sec:LG-FMO}In the forthcoming subsections, we consider the application
of the Leggett-Garg inequality to the FMO complex. We first discuss several
observables that one might measure in a Leggett-Garg protocol. We then obtain
analytical results when the dynamics are purely coherent. These analytical
results allow us to determine the time intervals between measurements in a
Leggett-Garg protocol that lead to the stongest violation of the inequality.
We finally use these time intervals in a numerical simulation of the
FMO\ dynamics that includes the effects of noise. The result is that several
observables exhibit a strong violation of the inequality for temperatures
below room temperature, and the violation persists in some cases up to room temperature.

\subsection{Observable for the Leggett-Garg Inequality in the FMO\ Complex}

Recall that the Leggett-Garg quantity in (\ref{eq:LG-quantity}) involves any
three dichotomic observables $Q\left(  t_{1}\right)  $, $Q\left(
t_{2}\right)  $, and $Q\left(  t_{3}\right)  $ measured at respective times
$t_{1}$, $t_{2}$, and $t_{3}$.

We have freedom in choosing both the observables that we measure and the times
at which we measure them. Perhaps the simplest dichotomic observable that we
can construct corresponds to the question \cite{P95}:

\begin{quote}
\textit{\textquotedblleft Is the system in state }$\left\vert \psi
\right\rangle $\textit{ or not?\textquotedblright}
\end{quote}

The two-element set of measurement operators\ for this question are as
follows:\ $\left\{  \left\vert \psi\right\rangle \left\langle \psi\right\vert
,I-\left\vert \psi\right\rangle \left\langle \psi\right\vert \right\}  $. We
assign the value $+1$ if the system is in the state $\left\vert \psi
\right\rangle $ and the value $-1$ otherwise. Let $Q_{\left\vert
\psi\right\rangle }$ denote the resulting observable where%
\begin{equation}
Q_{\left\vert \psi\right\rangle }\equiv\left\vert \psi\right\rangle
\left\langle \psi\right\vert -\left(  I-\left\vert \psi\right\rangle
\left\langle \psi\right\vert \right)  =2\left\vert \psi\right\rangle
\left\langle \psi\right\vert -I. \label{eq:LG-observable-psi}%
\end{equation}

We might build a dichotomic observable\ from states in the \textit{exciton
basis}. The exciton basis is the energy eigenbasis $\left\{  \left\vert
\phi_{m}\right\rangle \right\}  _{m=1}^{7}$\ of the Hamiltonian $H$ in
(\ref{eq:site-hamil}), where%
\[
\forall m\ \ \ \ \ \ H\left\vert \phi_{m}\right\rangle =E_{m}\left\vert
\phi_{m}\right\rangle .
\]
Then the dichotomic observable constructed from an energy eigenstate is
$Q_{\left\vert \phi_{m}\right\rangle }$. Note that observables of this form
commute with the Hamiltonian $H$\ in (\ref{eq:site-hamil}).

Another possibility is to build the dichotomic observable from the site basis.
This type of observable asks the question, \textit{Is the
excitation at site }$m$\textit{?}\ where $m\in\left\{
G,1,\ldots,7,S\right\}  $. The dichotomic observable constructed from a site
state is the \textit{site observable} $Q_{\left\vert m\right\rangle }$.
Observables of this form do not commute with the Hamiltonian $H$ in
(\ref{eq:site-hamil}).

\subsection{Analytical Results for Coherent Dynamics}%

\begin{figure*}
[ptb]
\begin{center}
\includegraphics[
height=3.8207in,
width=6.5864in
]%
{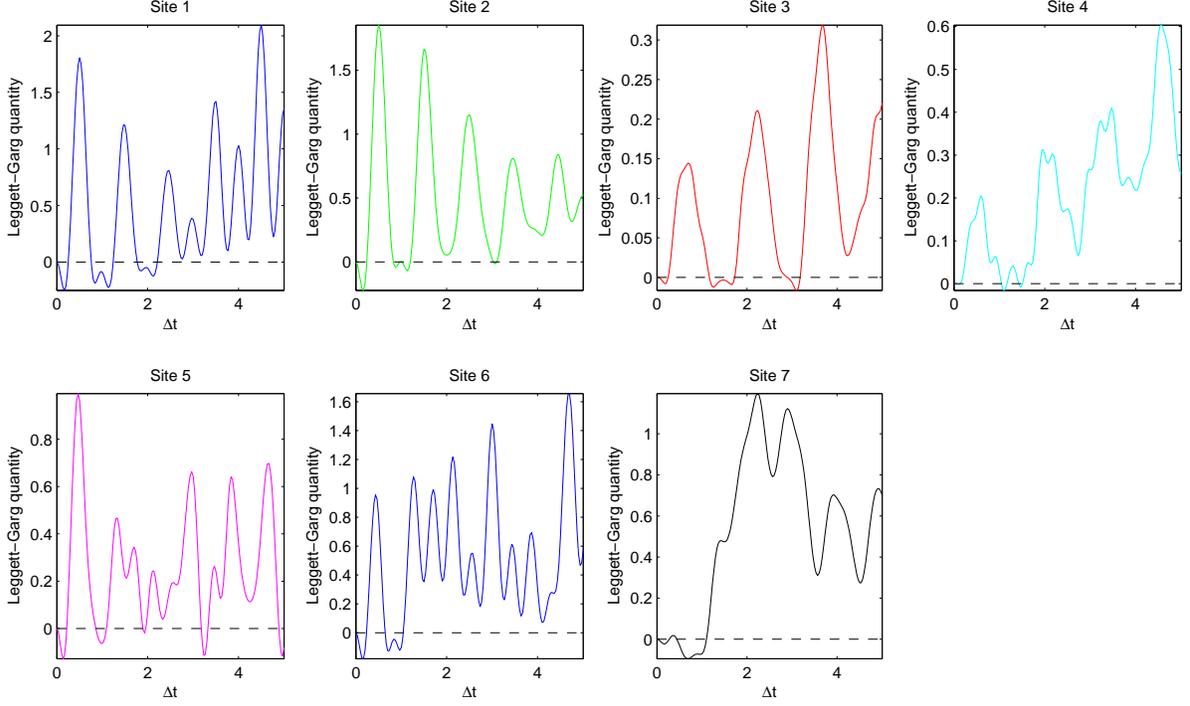}%
\caption{(Color online)\ The Leggett-Garg quantity as a function of the
uniform time interval $\Delta t$ for all seven sites in the FMO\ complex when
evolution is coherent and the initial state is the classical mixture
$\pi_{1,6}$. The units of $\Delta t$ are picoseconds. A dotted line divides
each plot into two regions. Points above a dotted line are in a
\textquotedblleft no violation\textquotedblright\ region, while points below
are in a \textquotedblleft violation\textquotedblright\ region. The convention
is the same in Figures~\ref{fig:sites-FMO-numerical},
\ref{fig:LG-violation-init-1}, and \ref{fig:LG-violation-site-6}. To show
overall behavior, the $\Delta t$ axis has a coarse scale up to five picoseconds (ps).
However, an experimental test would require control of $\Delta t$ to hundreds
of femtoseconds.}%
\label{fig:all-sites-coherent}%
\end{center}
\end{figure*}
Let us first suppose that evolution of the FMO\ complex is coherent and does
not include the noisy Lindblad evolution terms in (\ref{eq:lindblad-diss}%
-\ref{eq:lindblad-deph}). This assumption is unrealistic, but it gives a
starting point for understanding the Leggett-Garg inequality and the
FMO\ complex before proceeding with the full-blown evolution in
(\ref{eq:FMO-dynamics}).

The general form of the two-time correlation functions appearing in the
Leggett-Garg quantity $K$ in (\ref{eq:LG-quantity}) are as follows for a
coherent evolution:%
\begin{align*}
C_{1,2}  &  =\frac{1}{2}\text{Tr}\left[  Q\left(  t_{2}\right)  e^{-iH\left(
t_{2}-t_{1}\right)  }\left\{  Q\left(  t_{1}\right)  ,\rho_{t_{1}}\right\}
e^{iH\left(  t_{2}-t_{1}\right)  }\right]  ,\\
C_{2,3}  &  =\frac{1}{2}\text{Tr}\left[  Q\left(  t_{3}\right)  e^{-iH\left(
t_{3}-t_{2}\right)  }\left\{  Q\left(  t_{2}\right)  ,\rho_{t_{2}}\right\}
e^{iH\left(  t_{3}-t_{2}\right)  }\right]  ,\\
C_{1,3}  &  =\frac{1}{2}\text{Tr}\left[  Q\left(  t_{3}\right)  e^{-iH\left(
t_{3}-t_{1}\right)  }\left\{  Q\left(  t_{1}\right)  ,\rho_{t_{1}}\right\}
e^{iH\left(  t_{3}-t_{2}\right)  }\right]  ,
\end{align*}
where $H$ is the Hamiltonian in (\ref{eq:site-hamil}), $\rho_{t_{1}}$ is the
initial density operator of the FMO\ complex, $\rho_{t_{2}}\equiv
e^{-iH\left(  t_{2}-t_{1}\right)  }\rho_{t_{1}}e^{iH\left(  t_{2}%
-t_{1}\right)  }$, and $\left\{  \cdot,\cdot\right\}  $ is the anticommutator.
As indicated in Section~\ref{sec:LG}, the correlator $C_{1,3}$ characterizes
an experiment in which measurements occur at times $t_{1}$ and $t_{3}$ but no
measurement occurs at time $t_{2}$.

Suppose first that we prepare the FMO\ complex in some state $\left\vert
\psi\right\rangle $ and take the Leggett-Garg observable to be $Q_{\left\vert
\psi\right\rangle }\ $as in (\ref{eq:LG-observable-psi}). Thus, this
measurement is a \textquotedblleft survival probability\textquotedblright%
\ measurement \cite{P95,kofler:090403,Kof08}. Suppose further that we measure
$Q_{\left\vert \psi\right\rangle }$ at uniform time intervals $\Delta t$. A
straightforward calculation \cite{kofler:090403,Kof08} shows that the
Leggett-Garg quantity in (\ref{eq:LG-flip-2}) is equal to%
\begin{equation}
4\left\vert \left\langle \psi|\psi_{2\Delta t}\right\rangle \right\vert
^{2}-4\operatorname{Re}\left[  \left(  \left\langle \psi|\psi_{\Delta
t}\right\rangle \right)  ^{2}\left\langle \psi_{2\Delta t}|\psi\right\rangle
\right]  ,\label{eq:kofler-LG}%
\end{equation}
where $\left\vert \psi_{t}\right\rangle \equiv e^{-iHt}\left\vert
\psi\right\rangle $. Recall that a violation of the Leggett-Garg inequality
occurs when the above quantity drops below zero.%
\begin{table*}[tbp] \centering
\begin{tabular}
[c]{l|l|l}\hline\hline
$\rho_{t_{1}}$ & \textbf{Measurement} & \textbf{Leggett-Garg Quantity}%
\\\hline\hline
$\pi_{1,6}$ & $\left\vert 1\right\rangle $ & $2\left(  \left\vert \left\langle
1|6_{\Delta t}\right\rangle \right\vert ^{2}+\left\vert \left\langle
1|1_{2\Delta t}\right\rangle \right\vert ^{2}-\operatorname{Re}\left\{
\left\langle 1\right\vert \left(  \left\vert 1_{\Delta t}\right\rangle
\left\langle 1_{2\Delta t}\right\vert +\left\vert 6_{\Delta t}\right\rangle
\left\langle 6_{2\Delta t}\right\vert \right)  \left\vert 1\right\rangle
\left\langle 1|1_{\Delta t}\right\rangle \right\}  \right)  $\\\hline
$\pi_{1,6}$ & $\left\vert 6\right\rangle $ & $2\left(  \left\vert \left\langle
6|1_{\Delta t}\right\rangle \right\vert ^{2}+\left\vert \left\langle
6|6_{2\Delta t}\right\rangle \right\vert ^{2}-\operatorname{Re}\left\{
\left\langle 6\right\vert \left(  \left\vert 1_{\Delta t}\right\rangle
\left\langle 1_{2\Delta t}\right\vert +\left\vert 6_{\Delta t}\right\rangle
\left\langle 6_{2\Delta t}\right\vert \right)  \left\vert 6\right\rangle
\left\langle 6|6_{\Delta t}\right\rangle \right\}  \right)  $\\\hline
$\pi_{1,6}$ & $\left\vert 2\right\rangle $,\ $\ldots$, $\left\vert
5\right\rangle $, $\left\vert 7\right\rangle $ & $2\left(  \left\vert
\left\langle m|1_{\Delta t}\right\rangle \right\vert ^{2}+\left\vert
\left\langle m|6_{\Delta t}\right\rangle \right\vert ^{2}-\operatorname{Re}%
\left\{  \left\langle m\right\vert \left(  \left\vert 1_{\Delta t}%
\right\rangle \left\langle 1_{2\Delta t}\right\vert +\left\vert 6_{\Delta
t}\right\rangle \left\langle 6_{2\Delta t}\right\vert \right)  \left\vert
m\right\rangle \left\langle m|m_{\Delta t}\right\rangle \right\}  \right)
$\\\hline
$\left\vert 1\right\rangle \left\langle 1\right\vert $ & $\left\vert
1\right\rangle $ & $4\left\vert \left\langle 1|1_{2\Delta t}\right\rangle
\right\vert ^{2}-4\operatorname{Re}\left\{  \left\langle 1_{2\Delta
t}|1\right\rangle \left(  \left\langle 1|1_{\Delta t}\right\rangle \right)
^{2}\right\}  $\\\hline
$\left\vert 1\right\rangle \left\langle 1\right\vert $ & $\left\vert
2\right\rangle $,\ $\ldots$, $\left\vert 7\right\rangle $ & $2\left\vert
\left\langle m|1_{\Delta t}\right\rangle \right\vert ^{2}-4\operatorname{Re}%
\left\{  \left\langle m|1_{\Delta t}\right\rangle \left\langle 1_{2\Delta
t}|m\right\rangle \left\langle m|m_{\Delta t}\right\rangle \right\}
+2\operatorname{Re}\left\{  \left\langle 1_{2\Delta t}|m_{\Delta
t}\right\rangle \left\langle m|1_{\Delta t}\right\rangle \right\}  $\\\hline
$\left\vert 6\right\rangle \left\langle 6\right\vert $ & $\left\vert
6\right\rangle $ & $4\left\vert \left\langle 6|6_{2\Delta t}\right\rangle
\right\vert ^{2}-4\operatorname{Re}\left\{  \left\langle 6_{2\Delta
t}|6\right\rangle \left(  \left\langle 6|6_{\Delta t}\right\rangle \right)
^{2}\right\}  $\\\hline
$\left\vert 6\right\rangle \left\langle 6\right\vert $ & $\left\vert
1\right\rangle $, $\ldots$, $\left\vert 5\right\rangle $, $\left\vert
7\right\rangle $ & $2\left\vert \left\langle m|6_{\Delta t}\right\rangle
\right\vert ^{2}-4\operatorname{Re}\left\{  \left\langle m|6_{\Delta
t}\right\rangle \left\langle 6_{2\Delta t}|m\right\rangle \left\langle
m|m_{\Delta t}\right\rangle \right\}  +2\operatorname{Re}\left\{  \left\langle
6_{2\Delta t}|m_{\Delta t}\right\rangle \left\langle m|6_{\Delta
t}\right\rangle \right\}  $\\\hline
$\pi$ & all sites & $\frac{4}{d}-\frac{8}{d}\left\vert \left\langle
m|m_{\Delta t}\right\rangle \right\vert ^{2}+\frac{4}{d}\left\vert
\left\langle m|m_{2\Delta t}\right\rangle \right\vert ^{2}$\\\hline\hline
\end{tabular}
\caption{The first column lists the initial state of the FMO complex.
The second column lists the site that we measure in
a Leggett-Garg protocol. The third column gives the analytical form of the corresponding
Leggett-Garg quantity as a function of the uniform time interval $\Delta t$ when dynamics
are purely coherent. We use these formulas to
compute the results of Figure \ref{fig:all-sites-coherent} and Table \ref{tbl:coherent-times-2}.}\label{tbl:coherent-times}%
\end{table*}%

If the state $\left\vert \psi\right\rangle $ is an eigenstate $\left\vert
\phi_{m}\right\rangle $ in the exciton basis we should not expect to violate
the Leggett-Garg inequality because the observable $Q_{\left\vert \phi
_{m}\right\rangle }$ commutes with the Hamiltonian. We can confirm this
intuition by plugging the eigenstate $\left\vert \phi_{m}\right\rangle $ into
(\ref{eq:kofler-LG}). Doing so gives a value of zero for (\ref{eq:kofler-LG}),
thereby saturating the Leggett-Garg inequality, but yielding no violation.%
\begin{table}[tbp] \centering
\begin{tabular}
[c]{l|l|l|l}\hline\hline
\textbf{Initial State }$\rho_{0}$ & \textbf{Site} & $K$ & $\Delta t$
(ps)\\\hline\hline
$\pi_{1,6}$ & 1 & $-0.25053$ & $0.16678$\\\hline
$\pi_{1,6}$ & 2 & $-0.22321$ & $0.16678$\\\hline
$\pi_{1,6}$ & 3 & $-0.016389$ & $3.1021$\\\hline
$\pi_{1,6}$ & 4 & $-0.01574$ & $1.1008$\\\hline
$\pi_{1,6}$ & 5 & $-0.12782$ & $0.13343$\\\hline
$\pi_{1,6}$ & 6 & $-0.17994$ & $0.16678$\\\hline
$\pi_{1,6}$ & 7 & $-0.094719$ & $0.70048$\\\hline\hline
$\left\vert 1\right\rangle \left\langle 1\right\vert $ & 1 & $-0.4935$ &
$0.16678$\\\hline
$\left\vert 1\right\rangle \left\langle 1\right\vert $ & 2 & $-0.44335$ &
$0.16678$\\\hline
$\left\vert 1\right\rangle \left\langle 1\right\vert $ & 3 & $-0.065461$ &
$3.1355$\\\hline
$\left\vert 1\right\rangle \left\langle 1\right\vert $ & 4 & $-0.091838$ &
$1.7345$\\\hline
$\left\vert 1\right\rangle \left\langle 1\right\vert $ & 5 & $-0.08013$ &
$2.2015$\\\hline
$\left\vert 1\right\rangle \left\langle 1\right\vert $ & 6 & $-0.0097707$ &
$0.16678$\\\hline
$\left\vert 1\right\rangle \left\langle 1\right\vert $ & 7 & $-0.085607$ &
$1.034$\\\hline\hline
$\left\vert 6\right\rangle \left\langle 6\right\vert $ & 1 & $-0.0077476$ &
$0.13343$\\\hline
$\left\vert 6\right\rangle \left\langle 6\right\vert $ & 2 & $-0.0043891$ &
$1.034$\\\hline
$\left\vert 6\right\rangle \left\langle 6\right\vert $ & 3 & $-0.0032073$ &
$0.13343$\\\hline
$\left\vert 6\right\rangle \left\langle 6\right\vert $ & 4 & $-0.034082$ &
$1.4677$\\\hline
$\left\vert 6\right\rangle \left\langle 6\right\vert $ & 5 & $-0.27786$ &
$4.9701$\\\hline
$\left\vert 6\right\rangle \left\langle 6\right\vert $ & 6 & $-0.35011$ &
$0.16678$\\\hline
$\left\vert 6\right\rangle \left\langle 6\right\vert $ & 7 & $-0.18045$ &
$0.70048$\\\hline\hline
$\pi$ & all & 0 & all times\\\hline\hline
\end{tabular}
\caption{The first column lists the initial state of the FMO complex.
The second column lists the site observable that the Leggett-Garg protocol measures.
The third column lists the strongest violations for each site observable, and the fourth column gives
the corresponding time interval $\Delta t$ that leads to this violation.
(For comparison, the strongest possible violation of the inequality is $-0.5$.)
We obtained these values assuming that evolution is
coherent (though, we examined times up to $\Delta t = 5$ ps only). The last row
in the table indicates that we do not obtain a violation for any site observable
when the initial state is the maximally mixed state.}\label{tbl:coherent-times-2}%
\end{table}%

Thus, we set the Leggett-Garg observable to a site observable $Q_{\left\vert
m\right\rangle }$ where $m\in\left\{  1,\ldots,7\right\}  $. Sites 1 and 6 are
the chromophores that are closest to the chlorosome antenna and are thus most
likely to be the initial state of the FMO\ complex
\cite{1367-2630-11-3-033003}. The initial state $\rho_{t_{1}}$\ of the
FMO\ complex can be a pure state $\left\vert 1\right\rangle \left\langle
1\right\vert $ or $\left\vert 6\right\rangle \left\langle 6\right\vert $ or a
uniform classical mixture of sites $\left\vert 1\right\rangle $ and
$\left\vert 6\right\rangle $:%
\begin{equation}
\pi_{1,6}\equiv\frac{1}{2}\left(  \left\vert 1\right\rangle \left\langle
1\right\vert +\left\vert 6\right\rangle \left\langle 6\right\vert \right)  .
\label{eq:initial-state}%
\end{equation}
One might also consider setting the initial state to the maximally mixed state%
\[
\pi\equiv\frac{1}{7}\left(  \left\vert 1\right\rangle \left\langle
1\right\vert +\cdots+\left\vert 7\right\rangle \left\langle 7\right\vert
\right)  .
\]
Table~\ref{tbl:coherent-times} lists exact expressions for the Leggett-Garg
quantity in (\ref{eq:LG-flip-2}) for these different cases.

Figure~\ref{fig:all-sites-coherent} plots the Leggett-Garg quantity for each
site observable as a function of the uniform time interval $\Delta t$, when
the initial state is $\pi_{1,6}$. The result is that each of the seven site
measurements gives a violation of the Leggett-Garg inequality for some
intervals $\Delta t$\ when the dynamics are purely coherent. These results may
not be particularly surprising \cite{kofler:090403,Kof08} given that the
system is quantum, the measurements are sharp, and the dynamics are purely coherent.

Table~\ref{tbl:coherent-times-2}\ lists the time intervals that lead to the
strongest violation for $\Delta t\in\left[  0,5\right]  $ ps. We only consider
times up to five picoseconds (ps) because it is likely that the exciton will trap by this
time for the case of incoherent dynamics in the next section.
Table~\ref{tbl:coherent-times-2} also lists the time intervals that lead to
the strongest violation when the initial state is $\left\vert 1\right\rangle
\left\langle 1\right\vert $, $\left\vert 6\right\rangle \left\langle
6\right\vert $, or $\pi$.
We now use these time intervals for the more realistic case, where the
excitation in the FMO\ complex experiences the noisy contributions in
(\ref{eq:lindblad-diss}-\ref{eq:lindblad-deph}) from dissipation and dephasing.

\subsection{Numerical Results for Incoherent Dynamics}

\label{sec:numerical-sim}The Lindblad evolution in (\ref{eq:FMO-dynamics}) is
a more realistic model for evolution of the excitation in the FMO\ complex. It
incorporates the effects of excitonic recombination, trapping to the reaction
center, and enviromental dephasing noise, albeit in a Markovian context. The
two-time correlation functions appearing in the Leggett-Garg quantity $K$ in
(\ref{eq:LG-quantity}) must now be evaluated using the following more general
forms:%
\begin{align*}
C_{1,2}  &  =\frac{1}{2}\text{Tr}\left[  Q\left(  t_{2}\right)  \mathcal{N}%
_{t_{2},t_{1}}\left(  \left\{  Q\left(  t_{1}\right)  ,\rho\right\}  \right)
\right]  ,\\
C_{2,3}  &  =\frac{1}{2}\text{Tr}\left[  Q\left(  t_{3}\right)  \mathcal{N}%
_{t_{3},t_{2}}\left(  \left\{  Q\left(  t_{2}\right)  ,\mathcal{N}%
_{t_{2},t_{1}}\left(  \rho\right)  \right\}  \right)  \right]  ,\\
C_{1,3}  &  =\frac{1}{2}\text{Tr}\left[  Q\left(  t_{3}\right)  \mathcal{N}%
_{t_{3},t_{1}}\left(  \left\{  Q\left(  t_{1}\right)  ,\rho\right\}  \right)
\right]  ,
\end{align*}
where $\mathcal{N}$ is the superoperator that propagates the density operator
forward in time according to the evolution in (\ref{eq:FMO-dynamics}).

\begin{figure}
[ptb]
\begin{center}
\includegraphics[
height=2.9516in,
width=3.5405in
]%
{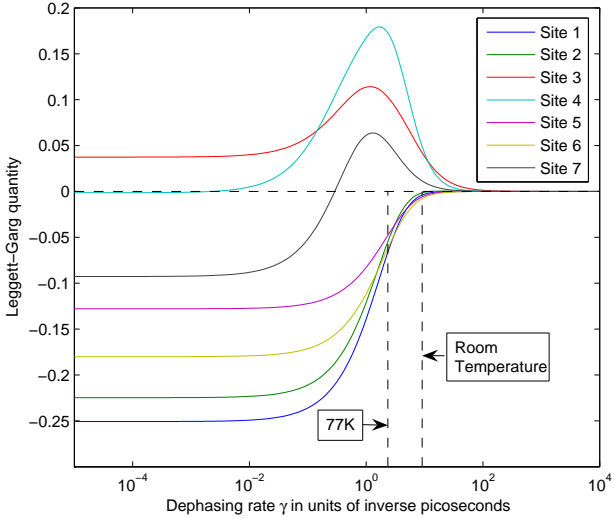}%
\caption{(Color online) The above figure displays the Leggett-Garg quantity as
a function of the dephasing parameter $\gamma$\ for each site observable
$Q_{\left\vert m\right\rangle }$, where $m\in\left\{  1,\ldots,7\right\}  $.
The site observables $Q_{\left\vert 3\right\rangle }$ and $Q_{\left\vert
4\right\rangle }$ do not give a violation of the Leggett-Garg inequality for
any amount of dephasing noise, even though we observed a violation in
Figure~\ref{fig:all-sites-coherent}\ with purely coherent dynamics. Site
observables $Q_{\left\vert 1\right\rangle }$, $Q_{\left\vert 2\right\rangle }%
$, $Q_{\left\vert 5\right\rangle }$, $Q_{\left\vert 6\right\rangle }$, and
$Q_{\left\vert 7\right\rangle }$ give a strong violation of the inequality for
low dephasing noise. Each of their corresponding Leggett-Garg quantities has a
smooth, monotonic transition to the \textquotedblleft no
violation\textquotedblright\ region for stronger dephasing noise, with the
Leggett-Garg quantity for site observables $Q_{\left\vert 1\right\rangle }$,
$Q_{\left\vert 2\right\rangle }$, $Q_{\left\vert 5\right\rangle }$,
$Q_{\left\vert 6\right\rangle }$ withstanding the strongest amount of
dephasing noise before they make a transition to the \textquotedblleft no
violation\textquotedblright\ region. Using the temperature analysis of the
environment in Ref.~\cite{1367-2630-11-3-033003}, a dephasing rate of 2.1
ps$^{-1}$ corresponds to a temperature of around 77${{}^\circ}$K and a
dephasing rate of 9.1 ps$^{-1}$ corresponds to a temperature of 298${{}^\circ}%
$K (room temperature). Site observables $Q_{\left\vert 1\right\rangle }$,
$Q_{\left\vert 2\right\rangle }$, $Q_{\left\vert 5\right\rangle }$, and
$Q_{\left\vert 6\right\rangle }$ give a violation for room temperature, with
the Leggett-Garg quantity respectively equal to $-0.0039$, $-0.0015$,
$-0.0059$, and $-0.0079$. It might be difficult for an experimentalist to
observe these violations at room temperature given that they are weak.}%
\label{fig:sites-FMO-numerical}%
\end{center}
\end{figure}
\begin{figure}
[ptb]
\begin{center}
\includegraphics[
height=2.6662in,
width=3.5405in
]%
{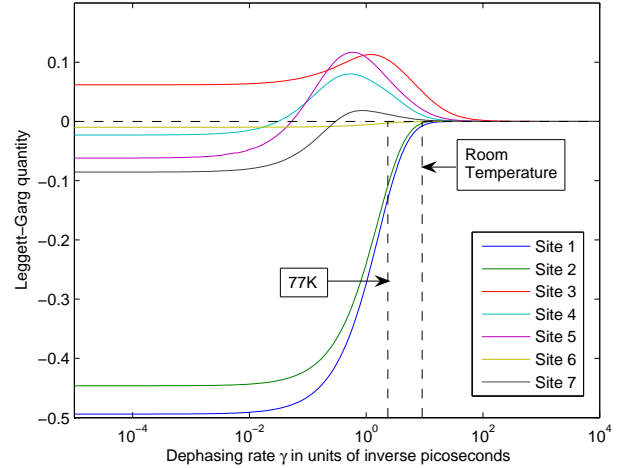}%
\caption{(Color online)\ This figure is similar to
Figure~\ref{fig:sites-FMO-numerical}, with the exception that the initial
state of the FMO\ complex is a pure state at site 1.~Site observables
$Q_{\left\vert 1\right\rangle }$ and $Q_{\left\vert 2\right\rangle }$ give a
violation for room temperature, with the Leggett-Garg quantity respectively
equal to $-0.0077$ and $-0.003$.}%
\label{fig:LG-violation-init-1}%
\end{center}
\end{figure}
\begin{figure}
[ptb]
\begin{center}
\includegraphics[
height=2.6662in,
width=3.5405in
]%
{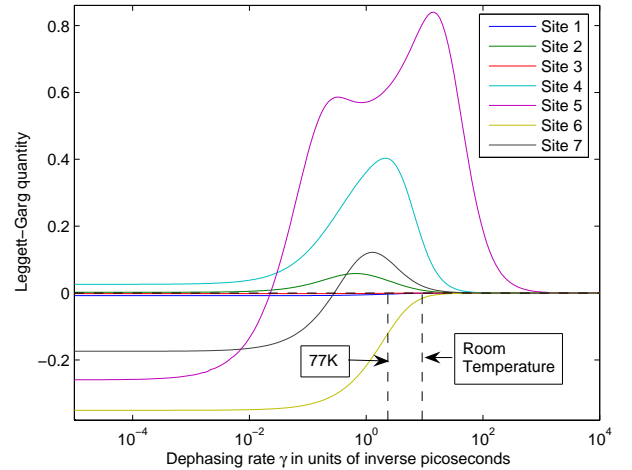}%
\caption{(Color online) This figure is similar to
Figure~\ref{fig:sites-FMO-numerical}, with the exception that the initial
state of the FMO\ complex is a pure state at site 6.~Site observable
$Q_{\left\vert 6\right\rangle }$ gives a violation for room temperature, with
the Leggett-Garg quantity equal to $-0.0155$.}%
\label{fig:LG-violation-site-6}%
\end{center}
\end{figure}

In order to perform a numerical simulation for the incoherent dynamics
(\ref{eq:FMO-dynamics}), we need to fix the trapping rate $\Gamma
_{\text{sink}}$, the recombination rates $\Gamma_{m}$, and the dephasing rates
$\gamma_{m}$. We take these rates from
Refs.~\cite{CCDHP09,1367-2630-10-11-113019}. Unless otherwise stated, the
trapping rate $\Gamma_{\text{sink}}=62.8/1.88$ cm$^{-1}$, corresponding to
about $2\pi\cdot c\cdot\Gamma_{\text{sink}}\approx6$ ps$^{-1}$, where $c$ is
the speed of light in units of cm$\cdot$ps$^{-1}$. The recombination rates
$\Gamma_{m}$ are uniform for all sites in the FMO\ complex so that $\Gamma
_{m}=\Gamma_{\text{recomb}}=1/\left(  2\cdot188\right)  $ cm$^{-1}$,
corresponding to about $2\pi\cdot c\cdot\Gamma_{\text{recomb}}\approx
5\times10^{-4}$ ps$^{-1}$. We assume that the dephasing rate is uniform so
that $\gamma_{m}=\gamma$ for all sites $m$.

As in the previous subsection, we choose the Leggett-Garg observable to be a
site observable $Q_{\left\vert m\right\rangle }$ where $m\in\left\{
1,\ldots,7\right\}  $. The initial state $\rho_{t_{1}}$\ can either be the
pure state $\left\vert 1\right\rangle \left\langle 1\right\vert $, $\left\vert
6\right\rangle \left\langle 6\right\vert $, or a uniform classical mixture of
sites $\left\vert 1\right\rangle $ and $\left\vert 6\right\rangle $ as in
(\ref{eq:initial-state}), for the same reasons mentioned in the previous
subsection (we do not consider the maximally mixed state because there is no
phenomenological evidence for this case, and furthermore, it does not give a
violation even in the coherent case). The time interval $\Delta t$ between
measurements of the Leggett-Garg observable is taken from
Table~\ref{tbl:coherent-times-2}. Figures~\ref{fig:sites-FMO-numerical},
\ref{fig:LG-violation-init-1}, and \ref{fig:LG-violation-site-6} display the
Leggett-Garg quantity as a function of the dephasing parameter $\gamma$\ for
each site observable $Q_{\left\vert m\right\rangle }$, where $m\in\left\{
1,\ldots,7\right\}  $.

The figures demonstrate that several site observables exhibit a violation even
as $\gamma$ increases. The temperature analysis of the environment in
Ref.~\cite{1367-2630-11-3-033003} indicates that a dephasing rate of 2.1
ps$^{-1}$ corresponds to a temperature of around 77${{}^{\circ}}$K and a
dephasing rate of 9.1 ps$^{-1}$ corresponds to a temperature of 298${{}%
^{\circ}}$K (room temperature). Under the assumption that this is
approximately correct, Figures~\ref{fig:sites-FMO-numerical},
\ref{fig:LG-violation-init-1}, and \ref{fig:LG-violation-site-6} predict a
violation at both temperatures. However, it might be difficult for an
experimentalist to observe these violations at room temperature given that
they are weak.

We have verified the robustness of these predictions against variations in the
dynamical parameters in (\ref{eq:FMO-dynamics}). Ref.~\cite{Adolphs20062778}%
\ mentions that the site energies of their calculated Hamiltonian are accurate
within $\pm2$ cm$^{-1}$. We therefore conducted several simulations that added
independent, zero-mean Gaussian noise with variance 2 to each entry in the
Hamiltonian matrix in (\ref{eq:site-hamil})\ to determine if the violations
would still hold under this slight perturbation. The result is that all room
temperature violations in Figures~\ref{fig:sites-FMO-numerical},
\ref{fig:LG-violation-init-1}, and \ref{fig:LG-violation-site-6} still hold,
and in fact, the values of the Leggett-Garg quantities are the same up to the
fourth decimal place. The trapping rates in the literature vary substantially,
including $0.25$ ps$^{-1}$ \cite{Adolphs20062778}, $1$ ps$^{-1}$
\cite{mohseni:174106}, or $4$ ps$^{-1}$ \cite{olaya-castro:085115}, so we have
checked our results for all of these choices. The plots for all these cases
are similar to Figures~\ref{fig:sites-FMO-numerical},
\ref{fig:LG-violation-init-1}, and \ref{fig:LG-violation-site-6} and have
approximately the same values for violations at room temperature. This finding
is reasonable given that most of our measurement times are less than the
average trapping times corresponding to these other trapping rates.

Our numerical simulations demonstrate that several choices of measurements
lead to a violation of the Leggett-Garg inequality even when noise processes
act on an excitation in the FMO\ complex. Experimental confirmation could
irrevocably exclude a class of macrorealistic theories from describing the
dynamics of the excitation.

\section{Conclusions}

We have framed tests that could be used to experimentally exclude a class of
macrorealistic theories, including a classical incoherent hoppping model, from
describing the room-temperature dynamical behavior of an excitation in the
FMO\ complex. To do so, we have introduced several examples of observables
that one might apply in a test of macrorealism, and our numerical simulations
predict that these observables lead to a violation of the Leggett-Garg inequality.

We discuss several ideas for furthering this research. The Leggett-Garg
measurements that we have considered here are projective measurements. In
practice, an experimentalist is unlikely to realize such an idealized
measurement. An experimentalist is more likely able only to effect a
\textquotedblleft noisy\textquotedblright\ or \textquotedblleft
fuzzy\textquotedblright\ measurement of the sites in the FMO\ complex. Along
these lines, Kofler \textit{et al}. have shown that \textquotedblleft
fuzzy\textquotedblright\ measurements may lead to our observation of a
classical world even in the presence of quantum
dynamics\cite{kofler:090403,Kof08}. Future work should formulate Leggett-Garg
inequalities involving quantities that are conveniently accessible to FMO
experimentalists (measurements in the site basis are currently infeasible
\cite{AR09}) and should consider the limits of experimental measurement
capabilities. Furthermore, future work should consider more realistic models
of noise in the FMO\ complex, potentially including correlated noise
\cite{N09,FNO09}\ and non-Markovian effects \cite{RCA09}.

One might consider an application of Bell's inequality to study the ability of
the FMO\ complex to preserve entanglement.

We have applied the Leggett-Garg inequality to the FMO complex, and it would
be valuable to apply the Leggett-Garg test to the dynamics of magnetoreception
in the avian compass \cite{K08a,K08b,K08c,CGB09}. It would also be interesting
to explore the Leggett-Garg inequality in artificial quantum networks with
particular noisy interactions. A study of this sort might lead to an increased
understanding of the dividing line between macrorealism and non-classicality
in more complicated systems. One might be able to apply the ideas in
Refs.~\cite{kofler:090403,Kof08,CCDHP09,1367-2630-10-11-113019}\ here.

\begin{acknowledgments}
The authors thank Andrew Cross, Alan Aspuru-Guzik, Kevin Obenland, Patrick
Rebentrost, and Leslie Vogt for useful discussions and thank Science
Applications International Corporation for supporting this scientific
research. MMW\ is especially grateful to Alan Aspuru-Guzik and Patrick
Rebentrost for their warm hospitality during his visit to Harvard and Leslie
Vogt for her suggestion to consider Leggett-Garg measurement time intervals
within the average trapping time of an exciton.
\end{acknowledgments}

\bibliographystyle{unsrt}
\bibliography{Ref}

\end{document}